%% file: 01-CC/0_Main.tex
\documentclass[sigconf,screen,nonacm]{acmart}

\usepackage[T1]{fontenc}
\usepackage{graphicx}
\usepackage{enumitem}
\usepackage{listings}

\AtBeginDocument{%
  }

\setcopyright{cc}
\setcctype[4.0]{by}
\acmYear{2026}
\acmConference[RiCE W1]{The First Reflection in Creative Experience (RiCE) Workshop}{July 13, 2026}{London, UK}
    
\begin{document}

\title{Meflex: Supporting Entrepreneurial Ideation Through Nonlinear Business Plan Writing}

\author{Lan Luo}
\email{lluo476@connect.hkust-gz.edu.cn}
\affiliation{%
  \institution{Hong Kong University of Science and Technology (Guangzhou)}
  \city{Guangzhou}
  \country{China}
}

\author{Dongyijie Primo Pan}
\email{dpan750@connect.hkust-gz.edu.cn}
\affiliation{%
  \institution{Hong Kong University of Science and Technology (Guangzhou)}
  \city{Guangzhou}
  \country{China}
}

\author{Junhua Zhu}
\email{junhuazhu@hkust-gz.edu.cn}
\affiliation{%
  \institution{Hong Kong University of Science and Technology (Guangzhou)}
  \city{Guangzhou}
  \country{China}
}

\author{Muzhi Zhou}
\email{mzzhou@hkust-gz.edu.cn}
\affiliation{%
  \institution{Hong Kong University of Science and Technology (Guangzhou)}
  \city{Guangzhou}
  \country{China}
}

\author{Pan Hui}
\email{panhui@hkust-gz.edu.cn}
\affiliation{%
  \institution{Hong Kong University of Science and Technology}
  \country{China}
}

\begin{abstract}
Traditional writing is often framed as a linear process of planning, drafting, revising, and finalizing. In entrepreneurship education, however, business plan (BP) writing is closely intertwined with ideation, as learners repeatedly generate, evaluate, abandon, and refine ideas while developing an opportunity. This nonlinear process places substantial cognitive demands on novice entrepreneurial students, yet existing creativity support tools rarely scaffold such higher-order thinking in BP writing. To address this gap, we present the Meflex System, a large language model (LLM)-based writing tool that integrates BP writing scaffolding with a nonlinear idea canvas to support iterative ideation through reflection and meta-reflection. We report findings from an exploratory user study with 30 participants examining the system’s usability and cognitive impact. Results show that Meflex effectively scaffolds BP writing, promotes divergent thinking through LLM-supported reflection, enhances meta-reflective awareness, and reduces cognitive load during complex idea development. These findings highlight the potential of nonlinear LLM-based writing tools to foster deeper thinking in entrepreneur education.
\end{abstract}


\maketitle
\textbf{Reference:}\newline
Lan LUO. 2026. Meflex: Supporting Entrepreneurial Ideation Through Nonlinear Business Plan Writing. In \textit{Proceedings of The First Reflection in Creative Experience (RiCE) Workshop (RiCE W1)}. ACM Creativity \& Cognition 2026, London, UK.

\section{INTRODUCTION}
\input{01-CC/1_Introduction}

\section{Literature Review}
\input{01-CC/2_RelatedWork.tex}

\section{Meflex System}
\input{01-CC/3_System}

\section{Method}
\input{01-CC/4_Method}

\section{Findings}
\input{01-CC/5_Findings}

\section{Discussion and Conclusion}
\input{01-CC/6_Discussion}

% \section{Conclusion}
% \input{01-CC/7_Conclusion}

% \section{Acknowledgment about the Use of LLM}
% The authors would like to acknowledge the use of the generative AI tool in this work. Specifically, GPT-4o by OpenAI was utilized to: (1) assist in language refinement, including grammar and style corrections of existing manuscript text, (2) generate LaTeX tables from the analyzed data results, and (3) GPT-4o API service was used in the system implementation. All interpretations, conclusions, and final content remain the responsibility of the authors.

\bibliographystyle{splncs04}
\bibliography{01-CC/Reference}

\section{Appendix}
\input{01-CC/8_Appendix}

\end{document}

%% file: 01-CC/1_Introduction.tex
% 非线性的文本写作
% 主要培养创新思维，通过发散和聚合思维。

\label{Introduction}
% BP对idea的作用：构建、评估、迭代。Gap：BP是线性写作; 初学者构建、评估、迭代idea有困难。
Business plan (BP) writing is a critical component of entrepreneurship education. While traditionally aimed at presenting business ideas to investors \cite{fayolle2015impact}, BP writing is increasingly seen in educational contexts as a pedagogical scaffolding that helps learners construct, evaluate, and iteratively refine their entrepreneurial ideas \cite{envick2020design,junhua2025designing}.
%Gap：BP写作还只是线性表达idea
Despite this pedagogical potential, BP writing remains a rigid, linear process centered on a single idea \cite{mansoori2020comparing,jackson2022advantage}. This stands in contrast to entrepreneurial ideation which is inherently dynamic and iterative \cite{ogutveren2019teaching}. It involves continuously revisiting initial abstract ideas through core business plan components such as problem identification, pain-point analysis and so on \cite{albers2019requirement}. 
%Gap：初学者构建、评估、迭代idea有认知困难
The substantial cognitive demands of iterative ideation often hinder idea development and opportunity recognition among novice entrepreneurial students \cite{greavu2019overview}.

% Reflection是ideation的策略
Reflection is a promising cognitive strategy for supporting entrepreneurial ideation through divergent thinking \cite{wardoyo2023determinant}. By prompting learners to critically examine their assumptions, consider alternative perspectives, and adapt their ideas in response to new insights, it fosters deeper cognitive engagement \cite{secundo2017activating,sadler1996learning}. In particular, it plays a vital role in fostering divergent thinking, which enables learners to generate creative possibilities \cite{razdorskaya2015reflection}, explore novel entrepreneurial directions \cite{bohlayer2025navigating}, and overcome cognitive fixation \cite{rosseel2022reflection}.
% Gap: idea也需要收束
However, effective ideation also demands the capacity to converge and refine ideas \cite{runco2010divergent}. Meta-reflection is a higher level of cognitive engagement and refers to the capacity to monitor, evaluate, and regulate one's own thinking \cite{thorpe2017co}. Through meta-reflection, learners gain awareness of their problem-solving approaches \cite{georgiev2011model}, recognize reasoning limitations \cite{thorpe2017co}, and adjust cognitive strategies accordingly \cite{zuber2015critical}. 
% 是writing tools，不是cirativity support tools
Yet few writing tools in entrepreneurial education are explicitly designed to scaffold these reflective and meta-reflective processes, leaving a critical gap in ideation support \cite{wardoyo2023determinant}.

% 是个LLM-based writing tools，不是creativity support tools
Recent advances in large language models (LLMs) offer new opportunities for addressing this gap. LLMs possess the capacity to engage in context-aware dialogue \cite{venkatraman2024collabstory}, generate alternative perspectives \cite{zhang2024comprehensive}, and prompt critical questioning \cite{reinhart2025llms}.
% Gap：现在tools集中在ideation，没有通过Bp writing来ideation的。
However, current LLM-based writing tools in entrepreneurial education primarily support linear and surface-level aspects of writing, such as grammar correction, content expansion, and structural coherence \cite{wambsganss2021arguetutor,ruan2021englishbot}, offering limited support for the iterative, non-linear nature of entrepreneurial ideation.
They also lack scaffolding for higher-order processes like reflection and meta-reflection, which demands both divergent and convergent thinking to explore possibilities, evaluate alternatives, and refine ideas over time \cite{kier2018entrepreneurial}.

To address this gap, we propose a non-linear writing tools with LLM support to facilitate reflection and meta-reflection, enabling learners to regulate, critique, and iteratively improve their thinking across multiple ideation cycles \cite{hacker1998metacognition,zimmerman1990self,greene2010fostering}. This leads to the central research question: \textit{How does the non-linear writing system with multi-agents scaffold BP writing and foster ideation?} To investigate it, we conducted an exploratory study with 30 participants using the Meflex system. Our main contributions are threefold. First, we developed \textbf{Meflex}, an multi-agent system that provides structured guidance and adaptive reflective scaffolding. Second, we introduced a non-linear writing interaction model that enables learners to flexibly revisit ideas through a visual canvas. Third, we empirically examined the influence of reflection and meta-reflection, revealing how these features enhance idea generation, synthesis, and cognitive integration in complex writing tasks.

% 结论
% The findings reveal that participants generally found Meflex usable and helpful in structuring their business plan writing; the system’s step-wise prompts and example-driven guidance effectively supported task completion, though some users required additional scaffolding to fully engage with the reflective features. Moreover, the integration of non-linear writing and LLM-assisted interactions fostered more divergent and flexible ideation, as participants reported that revisiting and rearranging ideas through the idea canvas helped them explore alternative directions and iteratively refine their concepts. Finally, GPT-generated reflective prompts and meta-reflection summaries enhanced users’ metacognitive awareness and supported synthesis across modules, facilitating deeper integration of ideas while also reducing cognitive load during complex planning tasks.

%% file: 01-CC/2_RelatedWork.tex
\label{RelatedWork}
\subsection{BP Writing as a Pedagogical Scaffold for Entrepreneurial Ideation}  
Business plans (BPs) generally provide a structured framework for articulating a new venture's products or services, funding sources, organizational structure, risks, and expected benefits \cite{cardamone2004business}. In entrepreneurship education, BP writing has become a common pedagogical activity, often associated with a prescriptive planning approach that emphasizes organizing resources, presenting implementation strategies, and conforming to institutional expectations \cite{honig2004entrepreneurship,karlsson2009judging,mansoori2020comparing}. However, this planning-oriented view can obscure the dynamic nature of entrepreneurial idea development, which often depends on external validation, real-world feedback, and adaptation to changing contexts \cite{wheadon2014business,neidell1988taking,baker1993business}. 

From this perspective, BP writing can be understood not only as documenting a venture concept, but also as a scaffold for clarifying assumptions, organizing fragmented ideas, and refining opportunities through feedback \cite{junhua2025designing,dasilva2014business}. Through iterative BP writing activities, students can externalize entrepreneurial thinking, respond to critique, and adapt their ideas as they learn from practice \cite{reagle2017idea,roy2021concept}. Building on this view, our work explores how BP writing can be transformed from a linear document-centered activity into a nonlinear structure that evolves with students' changing ideas.

\subsection{Reflection and Meta-reflection Interventions in Ideation}
Although BP writing can provide an external structure for entrepreneurial ideation, novice student entrepreneurs still need cognitive support to develop, evaluate, and revise their ideas. Many students struggle to move beyond initial concepts or explore alternative possibilities, highlighting the need to support not only the structural development of business ideas but also the cognitive processes that underlie creative ideation \cite{bock2018business}. 
Reflection supports this process by enabling learners examine and assess their current thinking, enabling iterative improvement and deeper cognitive engagement \cite{atkins1993reflection,secundo2017activating,sadler1996learning}.
In entrepreneurship education, reflective BP writing has been shown to support critical analysis and the iterative development of entrepreneurial ideas, especially when BPs are treated as active instruments for learning rather than merely as assessment tools \cite{jones2013moving,kier2018entrepreneurial}.

Beyond first-order reflection, meta-reflection helps learners examine how their ideas evolve over time and why key changes occur \cite{wang2025meta}. While reflection focuses on assessing current ideas, meta-reflection emphasizes the trajectory of thinking by surfacing how initial assumptions are challenged, what factors shape conceptual shifts, and how alternative directions emerge \cite{du2018meta,thorpe2017co}. This recursive awareness can strengthen cognitive flexibility and promote a sense of ownership and intentionality in the ideation process \cite{dozzi2024self}. In entrepreneurship education, such reflective and meta-reflective practices are especially valuable because opportunity development often involves nonlinear progress, iterative pivots, and uncertainty. Building on this view, our work combines reflection and meta-reflection within BP writing to help student entrepreneurs expand emerging ideas, evaluate alternatives, and make the rationale behind idea changes explicit during opportunity exploration.

\subsection{LLM-based Creativity Support for Ideation and Writing}
% LLM 已经能支持 idea generation / expansion；
% 但 ideation 不只是生成，还包括 evaluation / selection / refinement；
% writing 也不是简单产出文本，而包含 ideation、evaluation、revision 等认知过程；
% 因此，LLM tools 的关键不只是支持 ideation 或 writing 的某个结果，而是支持这些过程之间的转换；
% BP writing 正好需要这种支持，因为学生要在 idea development 和 structured writing 之间来回移动；
% gap：现有 BP writing tools 还没有充分支持 reflection/meta-reflection。
LLMs have shown potential in supporting individual ideation by increasing idea fluency, flexibility, and elaboration, helping users generate more ideas, explore more categories, and develop ideas with greater detail \cite{anderson2024homogenization,10.1145/3706598.3714034}. However, ideation involves not only divergent idea generation but also convergent processes of evaluating, selecting, and refining ideas for further development \cite{rodrigues2023creativity}. In parallel, LLM-based writing tools have been developed to provide feedback \cite{10.1145/3706598.3714316}, suggest revisions \cite{10.1145/3706598.3713691}, content coherence organization \cite{robidoux2007rhetorically}, and support domain-specific reasoning through structured prompts and iterative refinement \cite{lin2024rambler,10.1145/3613904.3642743}. Prior work on intelligent writing support further shows the value of designing such tools around the cognitive processes involved in writing rather than treating writing assistance as generic text generation \cite{goldi2024intelligent}.

For BP writing, these strands of work point to an important design opportunity. Novice entrepreneurial students should move between generating venture ideas, evaluating alternatives, organizing ideas into a structured plan, and revising them in response to feedback and uncertainty. Yet existing LLM-based creativity support tools for ideation \cite{anderson2024homogenization,10.1145/3635636.3656184} and writing \cite{siddiqui2025script,siddiqui2025script,https://doi.org/10.1111/bjet.13601} often support these activities separately, rather than scaffolding the reflective writing processes through which learners make sense of how ideas emerge, change, and connect over time. This motivates LLM-based BP writing tools that scaffold reflection and meta-reflection while preserving learners' active role in nonlinear entrepreneurial ideation.

%% file: 01-CC/3_System.tex
\label{UIdesign1}
\subsection{System Overview}

\begin{figure}[t]
  \vspace{-1em}
  \centering
  \includegraphics[width=\columnwidth]{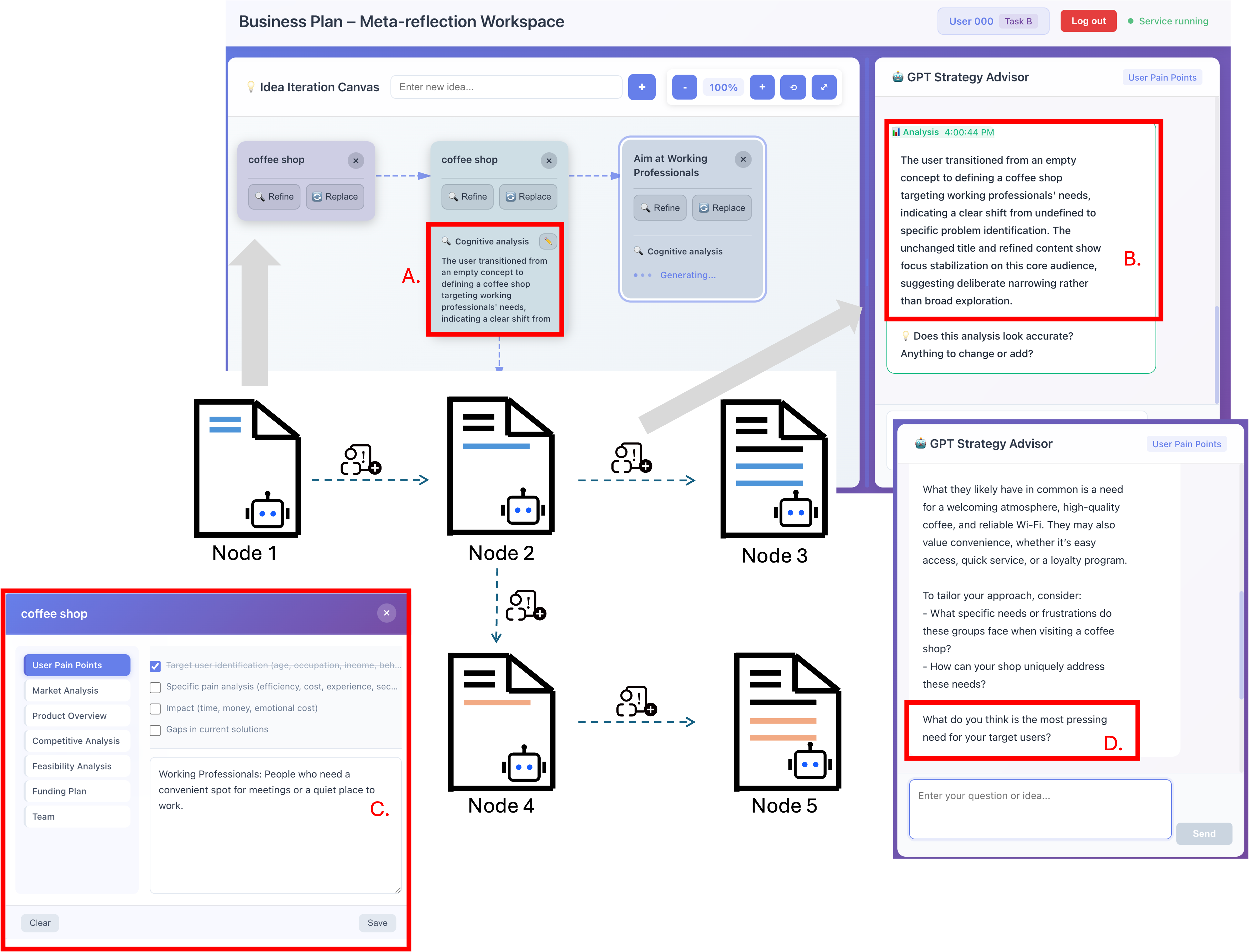}
  \caption{Three primary sections: the idea iteration canvas, the structural writing workspace, and the LLM assistant panel.}
  \label{fig:Interface}
  \vspace{-1em}
\end{figure}

The Meflex system interface in Fig.\ref{fig:Interface} is designed to support business plan writing through the scaffolding. The \textbf{ideation canvas} (Fig.~\ref{fig:Interface}, Left) allows users to progressively develop their entrepreneurial ideas through a visual workspace. Once an idea card is created, users can double-click it to open the \textbf{structure writing workspace} (Fig.~\ref{fig:Interface} (C)), which provides section-specific scaffolding for composing the business plan. This workspace offers predefined structural modules, each accompanied by LLM-assisted prompts and examples to guide writing in the \textbf{LLM assistant panel} (Fig.~\ref{fig:Interface}, Right). In parallel, each LLM response is followed by a reflective question (Fig.~\ref{fig:Interface} (D)) that encourages users to explore alternatives, or extend their ideas beyond the current formulation. 

The canvas is designed to support both divergent and convergent cognitive processes during BP ideation. To support divergence, users can extend idea cards horizontally or vertically. A horizontal extension represents incremental refinement, where the new card inherits the previous content and allows users to elaborate on the same idea. A vertical extension represents conceptual branching, where multiple cards develop from the same origin to explore alternative directions or variations. As ideas accumulate, the system supports convergence by helping users make sense of these branching paths. Each idea card includes an LLM-generated \textbf{meta-reflection} segment (Fig.~\ref{fig:Interface} (A)) that summarizes how the current idea evolved from earlier iterations. Users can further revise or deepen this meta-reflection through dialogue in the LLM panel. By linking idea branches with their underlying rationale, Meflex helps users compare alternatives, monitor conceptual shifts, and refine their venture direction during BP writing.

\subsection{Multi-Agents Structure and Prompts}
To support entrepreneurial ideation and business plan writing, we embedded five scaffolding strategies into the LLM-based system design \cite{junhua2025designing}: \textit{guidance} through structured steps, section-specific prompts, and analytic tools; \textit{contextualization} by situating ideas in meaningful entrepreneurial contexts; \textit{engagement} through interactive feedback \cite{clarizia2018chatbot}; \textit{interaction} through multi-turn dialogue for iterative clarification and revision \cite{tegos2015promoting}; and \textit{reflection} through prompts and summaries that help learners review assumptions, decisions, and writing progress \cite{sadler1996learning}. Together, these scaffolds guide learners from idea generation to business plan refinement.

Building on this design, the system uses a multi-agent LLM framework in which each major business plan section is supported by a dedicated agent. This structure enables section-specific guidance while maintaining coherence across the overall writing process. Each agent is implemented through role prompting and one-shot prompting: role prompting defines the agent's instructional focus, task boundary, interaction style, and section-level guidance, while one-shot prompting provides a concrete business plan example adapted from the U.S. Small Business Administration's template for \textit{We Can Do It Consulting} \cite{sba2023rebecca}. The example serves as a structural and stylistic reference, helping agents generate genre-aligned suggestions while allowing learners to adapt content to their own ideas. All LLM functions are powered by the DeepSeek API \cite{liu2024deepseek}. The full prompt is provided in Appendix~\ref{appendix:business-plan-agent-prompts}.

%% file: 01-CC/4_Method.tex
\label{Method}
We recruited 30 participants from the university, labeled P1 to P30, through an open call distributed via an online questionnaire. Participation was entirely voluntary, and the study aimed to assess the effectiveness of our system in supporting business plan writing and meta-reflection. The study protocol was reviewed and approved by the university's Institutional Review Board (IRB). Participants came from diverse academic backgrounds, including Communication and Media Arts, Artificial Intelligence, Human-Computer Interaction, Data Science and Analytics, Psychology, and Industrial Engineering. Regarding entrepreneurial experience, 21 participants reported no prior startup experience, while 9 reported some level of entrepreneurial experience.

The user study consisted of three phases. In the pre-study phase, participants completed a demographic questionnaire and received a 10-minute Zoom demonstration of the system's core features. During the in-study phase, participants selected one of ten predefined entrepreneurial topics, such as eco apps, travel planning, and e-bike sharing, and spent 40--60 minutes developing and refining their business ideas with the system. In the post-study phase, they completed a Likert-scale questionnaire on scaffolding and interaction design, followed by a 10--15 minute semi-structured interview. 

We analyzed the interview transcripts using thematic analysis. Two researchers independently coded the data, refined a shared codebook through discussion, and resolved disagreements in team meetings, identifying themes related to scaffolding support, reflection engagement, flexible writing structures, and idea development.

%% file: 01-CC/5_Findings.tex
\subsection{User Satisfaction with Scaffolding}
To examine participants' experiences with the LLM-supported writing scaffolds, we designed a 7-point Likert questionnaire based on the study goals and the key features of Meflex (1 = Strongly Disagree, 7 = Strongly Agree). These items were informed by prior work on scaffolding, including guidance, contextualization, engagement, interaction, and reflection \cite{junhua2025designing}. The full response distribution is provided in Appendix~\ref{appendix:satisfaction-results}.

Overall, participants reported positive experiences with the scaffolding features, with most items receiving mean scores between 5.5 and 6.5. The highest-scoring items (Q1, Q2, Q3, and Q10) suggest that participants completed required actions, invested time, and interacted with the interface repeatedly, while the slightly lower and more varied score for Q4 indicates differences in participants' interest in further developing their ideas. Qualitative feedback further showed that participants used the to-do list and LLM dialogue to organize their business plan writing (P2, P5, P8, P11, P14, P19, P25). The to-do list helped them break the task into manageable units and decide what to work on next (P2, P5, P14). As P8 noted, \textit{"The to-do list made me feel I had a roadmap... it didn't overwhelm me, and I could pick up right where I left off."} P7 similarly remarked, \textit{"It felt manageable because I could focus on one thing at a time. The list kept me on track."} These comments suggest that participants used the scaffolding structure to plan, resume, and sequence their writing actions.

Complementing the structured guidance, the LLM assistant provided adaptive support that accommodated users with varying levels of experience (P7, P12, P16, P21). Whether participants were familiar with business plan writing or not, the assistant offered tailored prompts and follow-up questions that adjusted to their current progress and input. For example, P21 with less experienced participant shared, \textit{"I didn’t know much about business plans, but the assistant broke things down for me and asked simple questions that helped me understand what to write."} Meanwhile, P16 with more confident user described: \textit{"I already had a basic plan in mind, and the assistant pushed me further—it kept asking questions that made me rethink my assumptions."} These examples show that Meflex supported both foundational writing guidance and deeper idea refinement, helping participants move between expressing business plan content and critically developing the ideas behind it.

\subsection{Reflection and Meta-reflection for Ideation}
Participants (P11, P18, P21) found GPT-generated reflection prompts valuable in encouraging deeper self-examination of their ideas. These prompts guided users to assess the validity, completeness, and consistency of their thinking, prompting them to question assumptions and consider alternative perspectives. As P26 noted, \textit{"The questions made me pause and think if my solution really solves the problem I described. I wouldn’t have done that on my own."} This kind of reflective scaffolding pushed participants beyond surface-level idea generation. P10 similarly shared, \textit{"It kept nudging me—‘have you thought about who really needs this?’ That helped me sharpen my target audience."} Rather than offering answers, the assistant acted as a metacognitive partner, supporting users in iteratively refining their ideas through structured self-questioning.

Beyond reflection, meta-reflection summaries also helped participants make sense of how different nodes and ideas were connected across their business plan drafts (P3, P12, P16). By synthesizing content across modules, these summaries offered a high-level overview that made the structure of their thinking more visible and easier to navigate. Several participants (P1, P3, P16) reported that the summaries encouraged them to revisit and refine earlier inputs based on new insights. For example, P15 noted, \textit{"After reading the summary, I realized my customer segment and value proposition didn’t quite match. That made me go back and rewrite part of it."} By surfacing potential inconsistencies, the system allowed users to focus their attention where revisions were needed, instead of manually cross-checking sections. Instead of manually tracing connections between sections, users could rely on the meta-reflection to highlight inconsistencies or gaps. P30 similarly reflected, \textit{"It reminded me that everything should connect—it wasn’t just writing section by section."} In this way, meta-reflections served as cognitive scaffolds that supported the integration of modular content into a coherent entrepreneurial narrative.

\subsection{Non-linear Cognitive Writing for Ideation}
The non-linear writing design in the Meflex system, particularly the ability to revisit, refine, and reorganize ideas across multiple stages, proved central to supporting entrepreneurial ideation. The system allowed users to navigate freely across modules and return to earlier thoughts, fostering a more dynamic and iterative creative process (P9, P11, P15, P25). Participants (P9, P25) reported that this modular organization helped them distinguish between ideas that were conceptually parallel and those that represented progressive refinement. P9 noted that separating early assumptions into different nodes clarified their relationships: \textit{"I could clearly see which parts were alternatives and which ones were building on each other. That made it easier to decide what to keep and what to revise"} . 

In addition to segmenting content, the node-based structure made the evolution of ideas more traceable (P8, P11, P14, P26). Participants could visually identify how one node branched into another, revealing how their thinking developed over time. This visibility prompted self-reflection and recursive refinement. As P14 noted, \textit{"When I looked back at how Node 4 came from Node 2, I realized my thinking had shifted—I could see that change clearly."} P11 echoed this, stating, \textit{"Seeing the path helped me understand what triggered each idea. It wasn’t random—I could trace the logic."} Similarly, P8 shared, \textit{"The branching view let me retrace my steps. I noticed how earlier thoughts shaped what came next, and sometimes that made me go back and revise."} P26 added, \textit{"It felt like watching my idea grow, not just listing points. I saw where things connected or didn’t, and fixed them."} Together, these reflections illustrate how the visualized structure of non-linear writing supported users in monitoring their ideation process and engaging in meaningful revision.

%% file: 01-CC/6_Discussion.tex
This study introduced Meflex, a human-AI co-writing system that supports BP writing in entrepreneurship education through nonlinear and scaffolded ideation. Existing LLM-based creativity tools often support either reflective writing or ideation, but rarely bring the two together to trace how entrepreneurial ideas evolve through iterative writing. Meflex addresses this limitation by embedding pedagogical scaffolding into a structured interface that integrates writing, reflective idea development, and meta-reflective synthesis.

Our evaluation shows that Meflex helps users deepen idea development, and maintain coherence across the overall business narrative. A key contribution of the system is its balance between structure and flexibility: users can focus on individual business plan sections while still seeing how each part contributes to the larger entrepreneurial logic. Overall, Meflex contributes to the design of LLM-driven educational writing systems by showing how human-AI collaboration can scaffold cognition process. We hope this work encourages future HCI research to further integrate reflection, meta-reflection into LLM-supported creativity tools.

%% file: 01-CC/8_Appendix.tex
\subsection{Prompts for Business Plan Agents}
\label{appendix:business-plan-agent-prompts}

The following prompt was used for the business plan agents. Each agent receives a role-specific task description and uses a one-shot example adapted from the U.S. Small Business Administration's sample business plan for \textit{We Can Do It Consulting} \cite{sba2023rebecca}.

\begin{lstlisting}[
basicstyle=\ttfamily\small,
breaklines=true,
frame=single,
columns=fullflexible
]
You are an entrepreneurial writing assistant supporting a learner in developing a business plan.

Your task is to provide section-specific guidance based on your assigned business plan role. You should help the learner clarify their ideas, identify missing information, improve the logic of the section, and connect their writing to realistic entrepreneurial contexts.

The system supports the following business plan agent roles:

{
  "User Pain Points": "Analyze and identify the target users' core pain points, their current solutions, and deeper unmet needs in real contexts.",

  "Market Analysis": "Discuss market size, growth trends, segmentation, and external drivers that shape the business landscape.",

  "Product Overview": "Define the product's value proposition, key features, use cases, and how it addresses user needs.",

  "Competitive Analysis": "Evaluate main competitors, alternatives, and differentiation strategy to build a sustainable advantage.",

  "Feasibility Analysis": "Assess technical, operational, and financial feasibility, including risks and implementation timeline.",

  "Funding Plan": "Design a realistic fundraising plan, including funding needs, investor profile, and return expectations.",

  "Team": "Present team strengths, member roles, relevant experience, and future hiring or organizational plans.",

  "Reflection": "Guide the user to reflect on their current reasoning and decisions by asking open-ended questions.",

  "Meta-Reflection": "Analyze the cognitive shift between idea versions and synthesize how the new thinking evolved."
}

One-shot example:

The following example is adapted from the SBA sample business plan for "We Can Do It Consulting." Use it as a reference for structure, specificity, and business-plan writing style. Do not copy the content directly.

Example role: Product Overview

Example business idea:
We Can Do It Consulting provides consulting services for small- and medium-sized companies.

Example section text:
We Can Do It Consulting provides consultation services to small- and medium-sized companies. The services include office management and business process reengineering to improve efficiency and reduce administrative costs. The company helps clients improve office processes such as job tracking, production, meetings, leadership, financial practices, and hiring practices. Its target customers include business owners, human resources directors, program managers, presidents, and CEOs of companies with 5 to 500 employees who want to increase productivity and reduce overhead costs.

What this example demonstrates:
1. The section begins with a clear description of the product or service.
2. It explains the practical value of the service.
3. It connects the service to specific customer needs.
4. It identifies the target users in concrete terms.
5. It uses a concise and formal business-plan style.

When responding, follow these steps:

1. Identify the learner's current business idea, section draft, or question.
2. Respond according to the assigned agent role.
3. Use the one-shot example as a structural and stylistic reference.
4. Provide concrete suggestions rather than general encouragement.
5. Ask follow-up questions when important information is missing.
6. When useful, provide a revised paragraph or partial draft for the current business plan section.

Input:
{
  "assigned_role": "%agent_role%",
  "learner_business_idea": "%business_idea%",
  "current_section_draft": "%section_draft%",
  "conversation_context": "%conversation_context%"
}

Output:
{
  "diagnosis": "Briefly summarize the current state of the learner's idea or draft.",
  "section_guidance": "Provide role-specific guidance for the assigned business plan module.",
  "revision_suggestions": [
    "Suggestion 1",
    "Suggestion 2",
    "Suggestion 3"
  ],
  "follow_up_questions": [
    "Question 1",
    "Question 2"
  ],
  "optional_revised_text": "If useful, provide a concise revised paragraph or partial draft."
}
\end{lstlisting}

\subsection{User Satisfaction with Scaffolding}
\label{appendix:satisfaction-results}

\begin{figure}[t]
    \centering
    \includegraphics[width=\columnwidth]{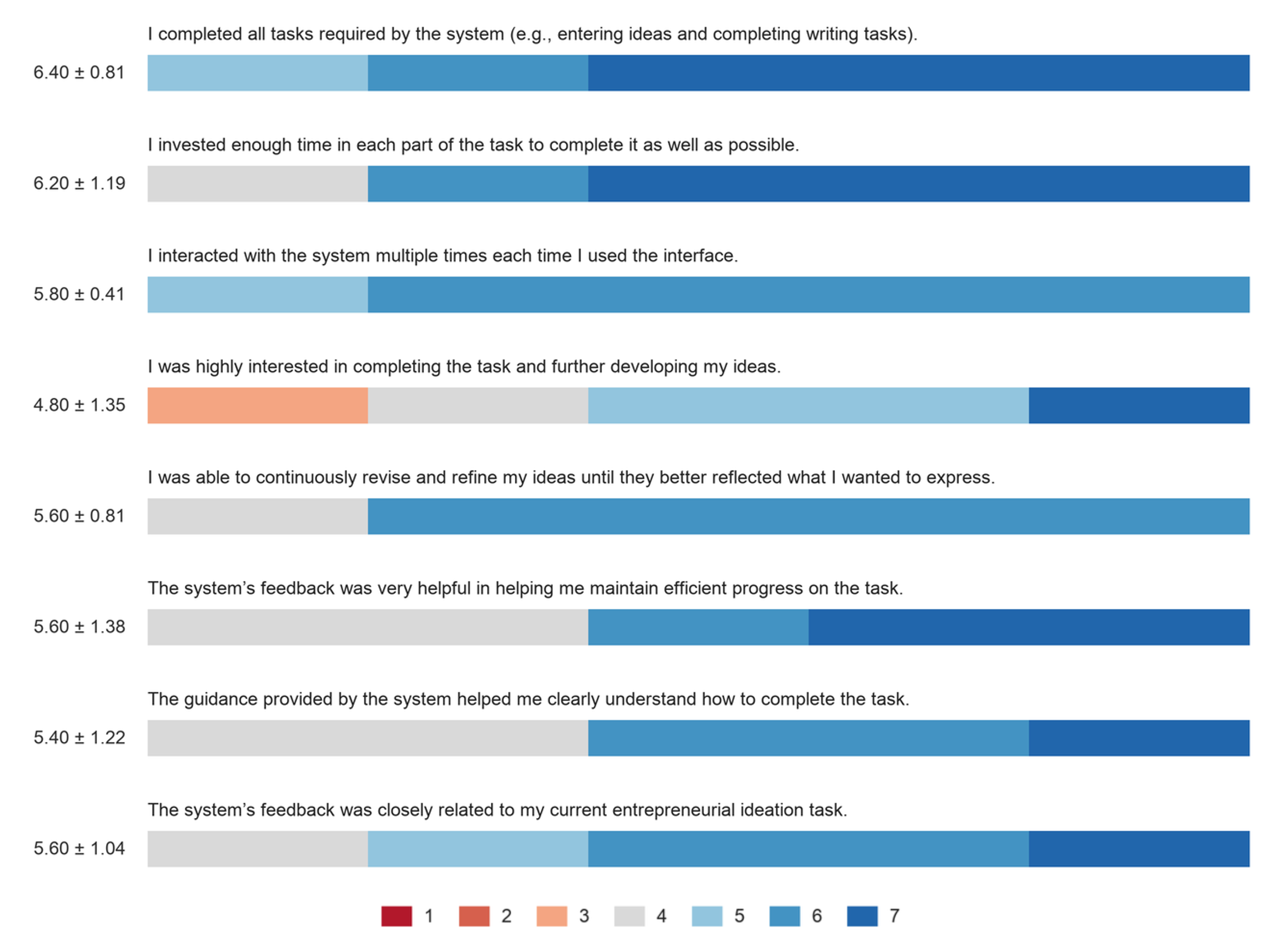}
    \caption{User Satisfaction with Scaffolding}
    \label{fig:ScaffodingSUS}
\end{figure}

Figure~\ref{fig:ScaffodingSUS} presents the full distribution of participants' responses to the user satisfaction questionnaire. Each stacked bar shows the proportion of ratings from 1 to 7 for one item, while the label on the left reports the mean score and standard deviation. These results complement the summary reported in the main text by showing item-level variation in participants' experiences with the scaffolding features.

